\newcommand{\CLASSINPUTtoptextmargin}{1.905cm}

\documentclass[conference, a4paper,10.9pt]{IEEEtran}
\IEEEoverridecommandlockouts
\usepackage{amsmath,amssymb,amsfonts}
\usepackage{algorithmic}
\usepackage{graphicx}
\usepackage{textcomp}
\usepackage{xcolor}
\def\BibTeX{{\rm B\kern-.05em{\sc i\kern-.025em b}\kern-.08em
    T\kern-.1667em\lower.7ex\hbox{E}\kern-.125emX}}
    
\usepackage[sorting=none]{biblatex}

\usepackage{balance}
\usepackage{csquotes}
\usepackage{comment}
\usepackage{amsmath}    
\usepackage{mathptmx}
\renewcommand{\figurename}{Fig.}
\usepackage[singlespacing]{setspace} 
\begin{document}

% Source https://www.physicsforums.com/threads/changing-the-order-of-references-using-bibtex-to-sequential.861217/
\title{SwiftSRGAN - Rethinking Super-Resolution for Efficient and Real-time Inference}

\author{\IEEEauthorblockN{Koushik Sivarama Krishnan}
\IEEEauthorblockA{Email: koushik.nov01@gmail.com}
\and
\IEEEauthorblockN{Karthik Sivarama Krishnan}
\IEEEauthorblockA{Email: ks7585@rit.edu}}

\maketitle

\begin{abstract}
In recent years, there have been several advancements in the task of image super-resolution using the state of the art Deep Learning-based architectures. Many super-resolution-based techniques previously published, require high-end and top-of-the-line Graphics Processing Unit (GPUs) to perform image super-resolution. With the increasing advancements in Deep Learning approaches, neural networks have become more and more compute hungry. We took a step back and, focused on creating a real-time efficient solution.  We present an architecture that is faster and smaller in terms of its memory footprint. The proposed architecture uses Depth-wise Separable Convolutions to extract features and, it performs on-par with other super-resolution GANs (Generative Adversarial Networks) while maintaining real-time inference and a low memory footprint. A real-time super-resolution enables streaming high resolution media content even under poor bandwidth conditions. While maintaining an efficient trade-off between the accuracy and latency, we are able to produce a comparable performance model which is one-eighth (1/8) the size of super-resolution GANs and computes 74 times faster than super-resolution GANs.\\
\end{abstract}


\section{Introduction}
\begin{figure}
    \includegraphics[width=0.48\textwidth, height=60mm]{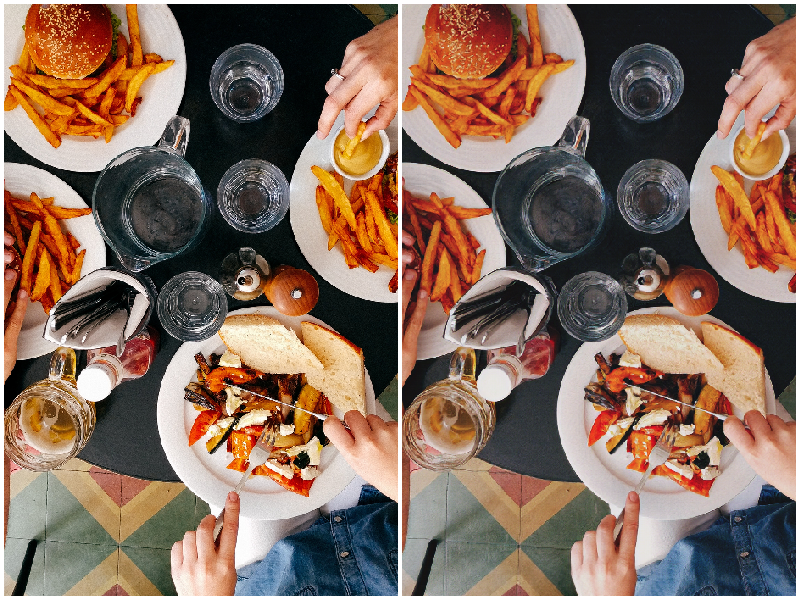}
    \caption{Image on the left is the original high-resolution image and the image on the right is the super-resolution image from SwiftSRGAN. The super-resolution(right) image is almost as good as the high-resolution image(left).}
    \label{fig:sample1}
    \hspace{0.05\textwidth}
    \end{figure}

 Image super-resolution has gained the attention of many researchers in the recent days, since the introduction of Convolutional Neural Networks for computer vision based tasks. Especially the reconstruction of a single high resolution image from a low quality low-resolution image has been a key focus among many research communities. Image super-resolution is the process of restoring and reconstructing the resolution of a noisy low-quality image into a very high quality high resolution image. For example, a reconstruction of a 256x256 pixels image into a 1024x1024 pixels image. As shown in Figure~\ref{fig:sample1}, this technique is very useful in various applications like up-scaling a low-resolution webcam image to perform facial recognition, up-sampling the medical images for identifying very small anomalies, reconstructing the world-war video feed into current super high resolution 4k frames, etc.
    
    With the recent advancements in deep learning, various newly proposed techniques and architectures like  attention\cite{DBLP:journals/corr/VaswaniSPUJGKP17}, transformers and GAN \cite{goodfellow2014generative} have become de-facto standard. There has been various updates and improvements to these proposed techniques since then and they have found their way into almost all the domains in deep learning. With the increasing advancements in Deep Learning approaches, there is also an increasing demand for high-end computing capabilities to perform computations of such complex neural network architectures\cite{8753848}. In today's real- world scenario, there is a very high demand for Mobile and embedded vision applications that performs in real-time. For applications such as Robotics and Wearable mobile devices, Augmented reality etc., the super-resolution tasks need to be performed in real-time with minimal latency and very low footprint requirements.
    
    This paper focuses on improving the latency of the Generative model by reducing the footprint and maintaining an efficient trade-off between the accuracy and the latency. Improving latency is benefited by reducing the computation size and introducing Depth-wise Separable Convolutions in place of standard Convolutions. This results in very few parameters required to be trained. 
    
    This technique has a significant application in many fields. When considering the  medical image super-resolution, Capturing a very high resolution Magnetic Resonance Imaging (MRI) are complicated when factors like scanning time, spatial convergence and signal-to-noise ratio are being involved. Hence applying the concepts of image super-resolution eliminates these problems by up-scaling the low-resolution image into high resolution images. The same case applies to X-rays \cite{9609375} and CT scans.
    
    Image super-resolution can also be applied to cloud gaming and media streaming services, where videos can be transmitted at a lower resolution and then up-scaled to a higher resolution on the edge. Streaming very high quality images often comes with a huge cost and high latency that will have an effect on the quality of the game-play. Image super-resolution can also be applied to surveillance and security cameras where the low-resolution footage can be up-scaled to a higher resolution for detecting  latent features present in the image frame that could be helpful in identifying suspect in real-time.
    
    Tasks like cloud gaming and media streaming happens in real-time, so we need a real-time and efficient solution that does not require top-of-the-line GPU to perform super-resolution. Cloud gaming and media streaming are typically processed on a data center outside the consumer's home and directly streamed to their at-home devices. Streaming very high-quality media content often comes with a huge cost and high latency that will affect the quality of the game-play. A real-time super-resolution approach would enable users to view high-resolution media content even on a poor internet connection without compromising on the clarity of the content. 
    
    Our approach also enables streaming graphics-intensive video games at low-resolution and then perform super-resolution on them without affecting the FPS(frames per second). We need an approach that works well even on low-end compute devices as top-of-the-line GPUs cannot be used everywhere. Our proposed approach can work even on low-end compute devices at 60 FPS.
    
\section{Related Work}
    Deep Learning has been growing exponentially in the past few years, with various newly proposed techniques and architectures like  attention, transformers\cite{DBLP:journals/corr/VaswaniSPUJGKP17} and GAN \cite{goodfellow2014generative} etc,. There has been various updates and improvements to these proposed techniques since then and they have found their way into almost all the domains in deep learning.
\subsection{Depth-wise Seperable Convolutions}
    Francois Chollet \cite{DBLP:journals/corr/Chollet16a} introduced Depth-wise Separable Convolutions. It is a two step operation: Depth-wise Convolution and Point-wise Convolution operations. The depth-wise Convolution is applied to a single image channel at a time, followed by the Point-wise Convolution, which is a 1x1 Convolution operation performed on M image channels. The standard Convolution operation both filters and combines input into a single step. The Depth-wise Convolution splits this into two steps, one for filtering and one for combining them. This results in very few parameters to tweak as compared to the normal Convolutions. They are also computationally cheaper and lighter on the disk.
    
    Since then, several other researchers have adapted this type of Convolution into their architectures. Howard et al. \cite{howard2017mobilenets} proposed a class of MobileNet architectures that are purely based on the Depth-wise Separable Convolutions. They added two new hyper-parameters, one for the width multiplier and another for the resolution multiplier. Width multiplier thins the network uniformly at each layer. For a selected layer with M channels and width multiplier $\alpha$, the number of input channels becomes $\alpha$M and output channels become  $\alpha$N. The second hyper-parameter resolution multiplier $\rho$, is used to reduce the dimensions of the input image.
\subsection{GANs}
    Ian et al. \cite{goodfellow2014generative} introduced the now, well known Generative models in 2014 that are trained simultaneously with discriminator model through adversarial process. The Generator tries to create fake images and the discriminator critics these images. As they progress through the training, the generator finally produces images that are indistinguishable from the real ones by the discriminator. Since then, many variations and updates has been made to this architecture for applications in different fields.

\subsection{CNN based approaches}
\subsubsection{\textbf{SRCNN}}
    Chao et al. \cite{DBLP:journals/corr/DongLHT15} introduced a Convolution based network architecture named as SRCNN that is structured into three stage process: Patch Extraction and Representation - this operation extracts overlapping patches from the low-resolution image and projects it into a high dimensional vector space. Non Linear Mapping - this operation non-linearly maps the resultant high dimensional vector from the previous operation onto another high dimensional vector. This results in a high resolution patch. Reconstruction - the generated high resolution patch is then aggregated to obtain the super-resolution output image. This proposed architecture achieves 32.52dB PSNR.
\subsubsection{\textbf{DRCN}}
    Jiwon et al. \cite{DBLP:journals/corr/KimLL15a} introduced a recursion based Convolution approach that used upto 16 recursions. And the main advantage of this approach is that, increasing the number of recursion can largely improve performance without introducing new parameters. They also observed that this approach is really hard to train with traditional gradient descent methods due to vanishing and exploding gradients. To tackle this problem, they introduced recursive-supervision and skip-connections to the network.
\subsubsection{\textbf{ESPCN}}
    Wenzhe et al. \cite{DBLP:journals/corr/ShiCHTABRW16} proposed a CNN architecture where the feature maps are extracted from the low-resolution images rather than the high-resolution images. They also proposed an efficient sub-pixel level Convolution operation that learns various up-sampling filters to up-scale the low-resolution feature maps into high resolution images. This also reduced the complexity of the super-resolution operation. 
    
    \begin{figure*}
    \includegraphics[width=\textwidth, height=8.5cm]{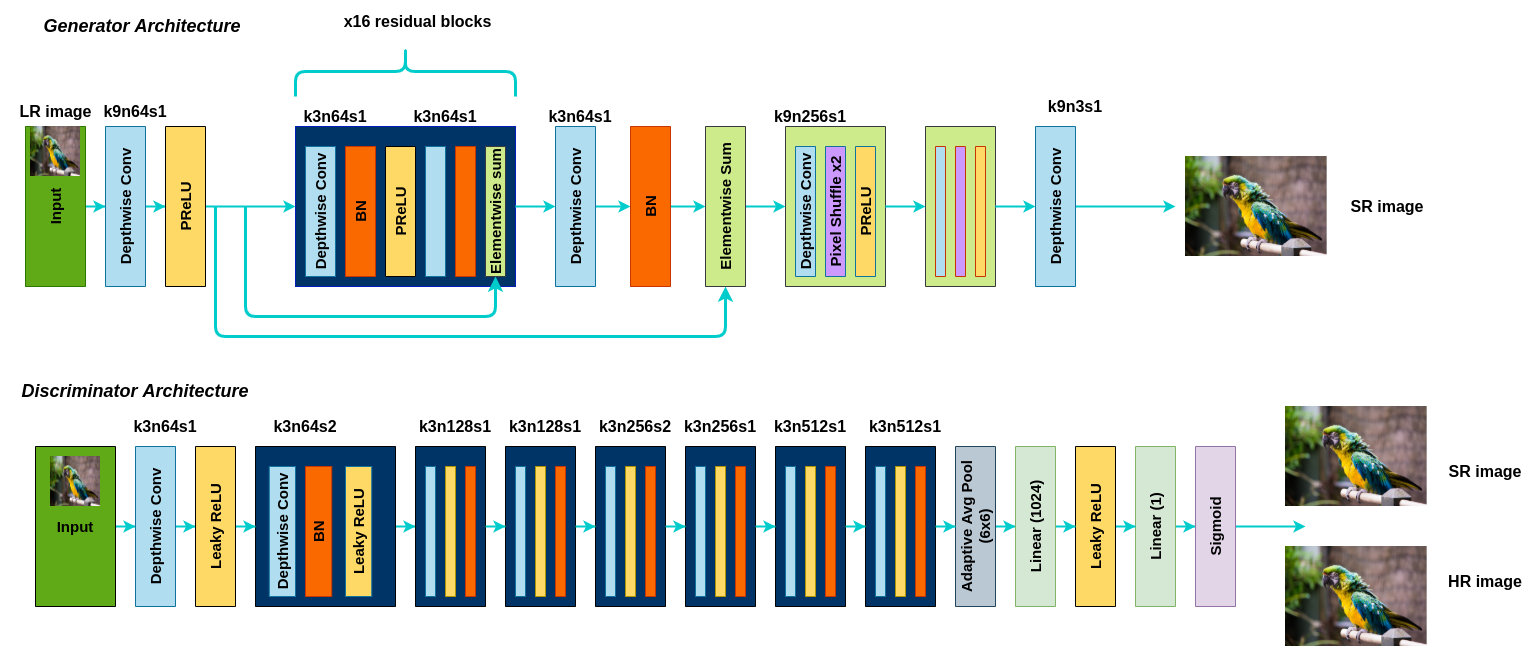}
    \caption{Architecture diagram of the Generator and Discriminator where 'k' is the kernel size, 'n' is the number of output channels, 's' is the stride.}
    \label{fig:architecture}
    \hspace{0.05\textwidth}
    \end{figure*}
    
\subsection{GAN based approaches}
\subsubsection{\textbf{SRGAN}}
    Christian et al. \cite{DBLP:journals/corr/LedigTHCATTWS16} introduced a Generative Adversarial Network-based architecture and proved this approach was better than using traditional methods previously proposed for image super-resolution. They introduced the Perceptual loss function based on the Mean Squared Error loss, which helped in removing the over-smoothing effects of MSE loss.
\subsubsection{\textbf{ESRGAN}}
    Xintao et al. \cite{DBLP:journals/corr/abs-1809-00219} modified the SRGAN\cite{DBLP:journals/corr/LedigTHCATTWS16} architecture by introducing Residual-in-Residual Dense Block for easier training. They have also improved the Discriminator architecture by using Relativistic Average GAN (RaGAN) \cite{DBLP:journals/corr/abs-1807-00734}, which learns to find the more realistic image rather than distinguishing the fake image.

\subsubsection{\textbf{DSCSRGAN}}
    Zetao et al. \cite{app10010375} proposed the use of Depth-wise Separable Convolutions in the intermediate layers and regular Convolutions in the first and last blocks. They used regular Convolutions for the discriminator and removed batch normalization from them. They also proposed a new loss function, frequency energy similarity loss function, which converts the spatial domain of the image into frequency domain. Then they calculate the similarity of frequency domain energies between the super-resolution image and the high-resolution target image.    

\section{Methodology}
\subsection{Dataset}
    To train this network, we merged two commonly available standard super-resolution datasets. DIV2K\cite{agustsson2017ntire} \cite{timofte2017ntire}  dataset and the Flickr2K \cite{lim2017enhanced} dataset. DIV2K\cite{agustsson2017ntire} \cite{timofte2017ntire} contains 800 high-resolution images for training, 100 for validation, and 100 for testing. The Flickr2K \cite{lim2017enhanced} dataset has 2650 high-resolution images collected from the Flickr website. We used Set5\cite{Huang-CVPR-2015}  and Set14\cite{Huang-CVPR-2015} test sets for evaluating our model. Set5\cite{Huang-CVPR-2015} test set has 5 high-resolution images and Set14\cite{Huang-CVPR-2015} has 14 high-resolution images in it.
    
     \begin{table}
    \caption{Dataset Split} 
    \renewcommand*\arraystretch{1.1}
    \normalsize
   \noindent \begin{tabular}{|p{20mm}||p{55mm}|}
    
    \hline
    Split & Number of High-Resolution images\\
    \hline
    Train    & 800 (DIV2K) + 2650 (Flickr2K) \\
    Validation  & 100 \\
    Test  & 100 + 5 (Set5) + 14 (Set14) \\
     \hline
    \textbf{Total}   & \textbf{3669} \\
    \hline
    \end{tabular}
    \end{table}

\subsection{Dataset Preprocessing}
    The high resolution images are augmented using the albumentations \cite{info11020125} library. This helps in making distinctive images that helps the deep learning model to learn much better and prevent overfitting. The high-resolution images are randomly cropped into 1024x1024 images. A random horizontal flip and random rotation to 90° is performed on these cropped images. This is then down-sampled using the bicubic operation to get low-resolution 256x256 images.

\subsection{Network Architecture}
    The generator architecture as shown in Figure~\ref{fig:architecture} consists of Depth-wise Separable Convolutions followed by Batch Normalization and PReLU as the activation function. Since our main goal is to perform real-time inference even on low-end GPUs, we have replaced regular Convolutions with Depth-wise Separable Convolutions. This has shown a significant reduction in inference time. The generator consists of 16 residual blocks with Depth-wise Convolution, followed by batch normalization and PReLU as activation function. This is followed by another Depth-wise Convolution, batch normalization, and lastly, an element-wise sum operation to sum the previous block's output and current output. We then up-sample the image by passing it through the up-sample Block twice. The up-sample Block contains Depth-wise Separable Convolutions followed by two Pixel Shuffle layers and PReLU as the activation function. This is then passed into the final Depth-wise Separable Convolution layer with a kernel size of 9, number of output channels as 3, and a stride of 1.
    
    The design is based on standard SRGAN \cite{DBLP:journals/corr/LedigTHCATTWS16} architecture, and expands by the introduction of Depth-wise Separable Convolutions which benefits for faster inference and parameter efficiency. Given an input low-resolution image of dimensions 3x256x256, the generator outputs a 3x1024x1024 super-resolution image instantaneously.
    
    The discriminator architecture consists of 8 Depth-wise Separable Convolution blocks. All blocks have a Depth-wise Separable Convolution, followed by batch normalization and LeakyReLU (negative slope=0.2) as activation function. The first block does not have a batch normalization layer. The output of the last Convolution block is passed to the adaptive average pooling layer that gives an output size of 6x6. The output of the adaptive average pooling layer is flattened and passed into a linear layer of 1024 neurons. The main objective of the discriminator is to classify super-resolution images as fake and high-resolution images as real.
    
\subsection{Perceptual Loss Function with MobileNet}
    The perceptual loss function was introduced by Christian et al. \cite{DBLP:journals/corr/LedigTHCATTWS16} which is based on the MSE loss. As our goal was to reduce both training and inference time, We improved on this loss function by replacing VGG \cite{simonyan2015deep} network with MobileNetV2 \cite{DBLP:journals/corr/abs-1801-04381} for efficient memory usage and training time. Since the definition of the perceptual loss function is very crucial for the performance of the generator, we only tweaked the network used. The perceptual loss function is formulated as the weighted sum of adversarial loss and, the content loss as:
    \begin{equation}
    \underbrace{l^{SR}}_\text{Perceptual loss} = \underbrace{l_X^{SR}}_\text{content loss} + \underbrace{1e-3l_{Gen}^{SR}}_\text{adversarial loss} 
    \end{equation}

\subsubsection{\textbf{Content Loss}}
    We build upon the proposed content Loss by Christian et al. \cite{DBLP:journals/corr/LedigTHCATTWS16} by replacing VGG19 \cite{simonyan2015deep} with MobileNetV2 \cite{DBLP:journals/corr/abs-1801-04381}. We define this MobileNet Loss based on the ReLU6 activation layers of the pre-trained network. \(\phi_{i}\) denotes the output of i-th Convolution block after activation. We used the output of the 16 th block for our loss function i.e. i = 16. This MobileNet loss is defined as the euclidean distance between the target image \( I^{HR}\) and the output feature map of the MobileNet loss \(G_{\theta_G}(I^{LR})\).
    
    \begin{equation}
        l_{MobileNet}^{SR} = \frac{1}{W_{i}H{i}} \Sigma_{x=1}^{W_i} \Sigma_{y=1}^{H_i}(\phi_{i}(I^{HR})_{x,y} - \phi_{i}(G_{\theta_G}(I^{LR}))_{x,y})^2
    \end{equation}
    
    Here $W_i$ and $H_i$ are feature map dimensions within the MobileNetV2 \cite{DBLP:journals/corr/abs-1801-04381} network.
    
\subsubsection{\textbf{Adversarial Loss}}
    We also add adversarial loss to our content loss. This encourages our network to produce images that are close to natural images, by trying to fool the discriminator. This generative loss is formulated as:
    \begin{equation}
        l_{Generator}^{SR} = \Sigma_{n=1}^{N} - \log D_{\theta_D}(G_{\theta_G}(I^LR))
    \end{equation}
    Where $D_{\theta_D}(G_{\theta_G}(I^LR))$ is the probability that the super-resolution image is a natural High-resolution image. And, instead of reducing $[1 - D_{\theta_D}(G_{\theta_G}(I^LR))]$, we minimize $- \log D_{\theta_D}(G_{\theta_G}(I^LR))$ for better gradients \cite{NIPS2014_5ca3e9b1}.

\section{Experiments and Analysis}
\subsection{Experimental Setup}
   We trained our Swift-SRGAN on Tesla P100 GPU with high-resolution images of dimensions 1024x1024x3 and, low-resolution images of dimensions 256x256x3. We used the AdamW \cite{DBLP:journals/corr/abs-1711-05101} optimizer with ReduceLROnPlateau learning rate scheduler and a batch size of 32. We also used the PyTorch's mixed-precision support to train our models. The low-resolution images are passed into the generator to get the super-resolution image. This is then passed into the perceptual loss function and the generator loss is calculated. This perceptual loss is simply adversarial loss + content loss. Then both high resolution and super-resolution images are passed into the discriminator to calculate the discriminator loss.

\subsection{Evaluation Criteria}
    In this study, we used Peak-Signal-to-Noise-Ratio(PSNR) and Structural Similarity Index Measure (SSIM)  as scoring metrics. PSNR is simply a ratio between maximum power of image to power of noise that affects the image quality. The higher the PSNR, better the super-resolution output image when compared to the high resolution target image.
    \begin{equation}
    PSNR  = 10 \cdot \log_{10} (\frac{MAX_1^2}{MSE})
    \end{equation}
    \begin{equation}
          = 20 \cdot \log_{10} (\frac{MAX_1}{MSE})
    \end{equation}
    \begin{equation}
          = 20 \cdot \log_{10} (MAX_1) - 10 \cdot \log_{10} (MSE)
    \end{equation}

    Structural Similarity Index Measure(SSIM) is a perceptual image metric that quantifies image degradation between the two images. It is calculated based on 3 comparison measurements: structure, luminance and contrast. This metric lies between 0.0 to 1.0 where 1.0 signifies perfect replica of the high resolution image.

    Luminance comparison function,
    \begin{equation}
        l(x, y) = \frac{2 \mu_x \mu_y + c_1}{\mu_x^2 \mu_y^2 + c_1}
    \end{equation} 
    Contrast comparison function,
    \begin{equation}
        l(x, y) = \frac{2 \sigma_x \sigma_y + c_2}{\sigma_x^2 \sigma_y^2 + c_2}
    \end{equation}
    Structure comparison function,
    \begin{equation}
        l(x, y) = \frac{2 \sigma_{xy} + c_3}{\sigma_x \sigma_y + c_3}
    \end{equation}
    and, 
    \begin{equation}
        c3 = c_2/2
    \end{equation}
    SSIM is the weighted combination of these measures,
    \begin{equation}
        SSIM(x, y) = [l(x, y)^\alpha \cdot c(x, y)^\beta \cdot s(x, y)^\gamma]
    \end{equation}

\section{Results}
    \begin{figure}
    \includegraphics[width=0.480\textwidth, height=60mm]{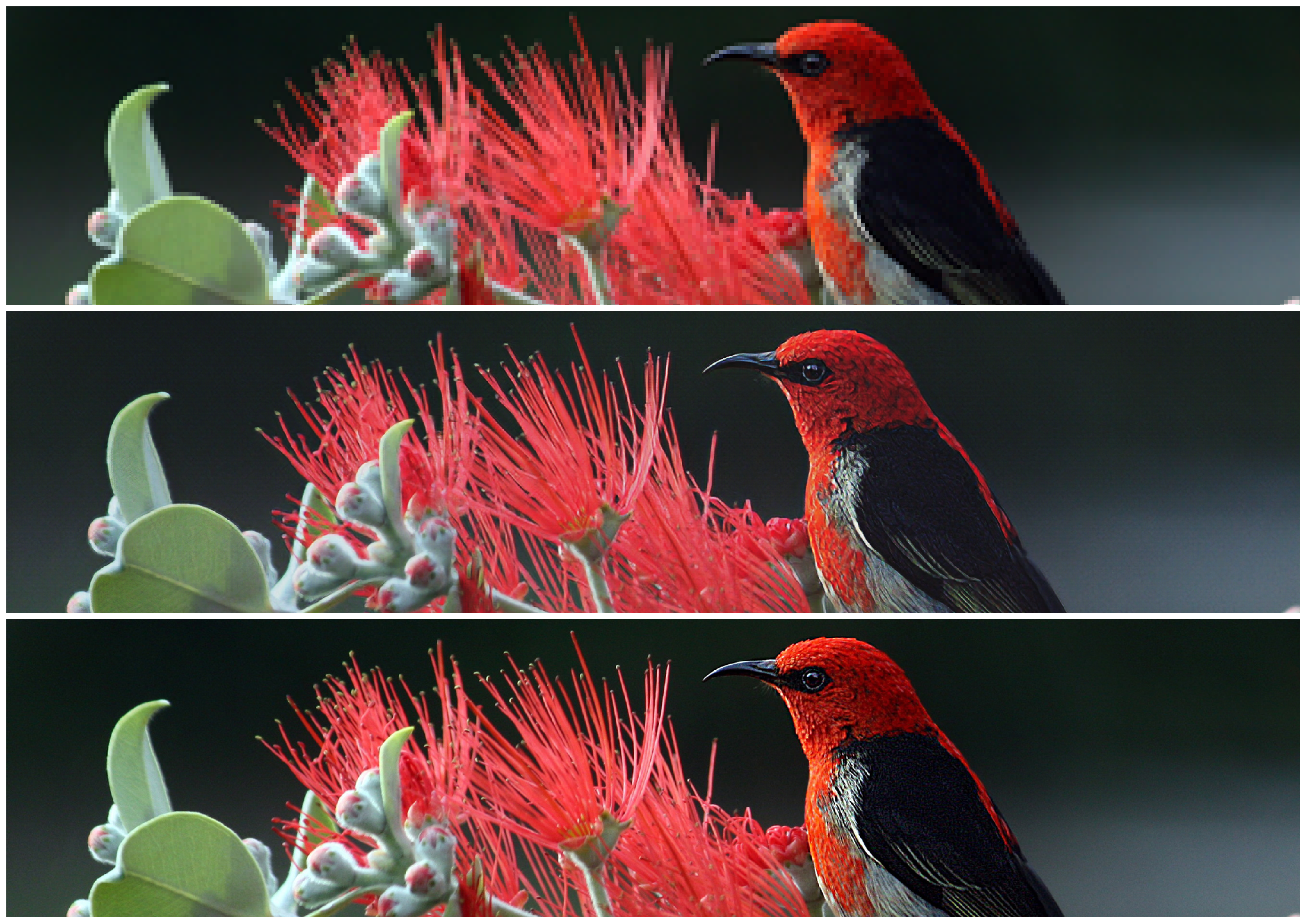}
    \caption{Low-resolution Bicubic image (top), Super-resolution image (middle) and, High-resolution image (bottom). As you can see from the images, our model has reproduced all the lighting, reflections (in the eye) and texture details seen on the high resolution image.}
    \label{fig:bird}
    \hspace{0.05\textwidth}
    \end{figure}
    
    \begin{table}
    \caption{Performance Analysis} 
    \renewcommand*\arraystretch{1.15}
    \normalsize

    \centering\begin{tabular}{|p{10mm}|p{35mm}|p{10mm}|p{10mm}|}
    
    \hline
    Test Set  & Model & PSNR & SSIM  \\
    \hline
    Set5 & ESRGAN & 32.7 & 0.9011  \\
    Set5 & SRGAN & 29.40 & 0.8501  \\
    Set5 & \textbf{Swift-SRGAN}(ours)  & 25.13 & 0.7940\\
    \hline
    Set5 & ESRGAN & 28.7 & 0.7917  \\
    Set14 & SRGAN &  26.02 & 0.7397 \\
    Set14 & \textbf{Swift-SRGAN}(ours) & 23.29 & 0.7012\\
    \hline
    \end{tabular}
    \end{table}
    
    \begin{table}
    \caption{Comparison of Inference time between SRGAN and Swift-SRGAN}
    \renewcommand*\arraystretch{1.15}
    \normalsize
    \centering\begin{tabular}{|p{22mm}|p{25mm}|p{25mm}|}
    \hline
   Upsampling Resolution  & Model & Inference time per-frame (in ms)\\
    \hline
    270p to 1080p & SRGAN & 812\\
    270p to 1080p & ESRGAN & 974\\
    270p to 1080p & \textbf{Swift-SRGAN}(ours)  & \textbf{5.605}\\
    \hline
    540p to 4K & SRGAN & 1210\\
    540p to 4K & ESRGAN & 1330\\
    540p to 4K & \textbf{Swift-SRGAN}(ours) & \textbf{16.18}\\
    \hline
    \end{tabular}

    \end{table}
    
To evaluate the performance of our model, we compared the inference time of our proposed approach to the inference time of the  SRGAN\cite{DBLP:journals/corr/LedigTHCATTWS16} architecture on Tesla K80 GPU. From the table II, we can clearly see that our proposed architecture has outperformed SRGAN\cite{DBLP:journals/corr/LedigTHCATTWS16}  and ESRGAN\cite{DBLP:journals/corr/abs-1809-00219} by a large magnitude in inference time. The inference time is calculated by the amount of time it takes to upsample a single image frame (in ms). Our proposed architecture achieves a  swift inference time of about 5.605 ms to up-sample a single image while the SRGAN\cite{DBLP:journals/corr/LedigTHCATTWS16} took 812 milliseconds to up-sample the same. This result signifies that our proposed architecture can up-sample 100 frames a second which is upto a 100 times faster than SRGAN\cite{DBLP:journals/corr/LedigTHCATTWS16} and ESRGAN\cite{DBLP:journals/corr/abs-1809-00219}. Hence our proposed method achieves an efficient up-sampling  of videos and images on the fly.

We also compared the PSNR and the SSIM scores of ESRGAN\cite{DBLP:journals/corr/abs-1809-00219}, SRGAN\cite{DBLP:journals/corr/LedigTHCATTWS16} and Swift-SRGAN(ours) on the standard benchmark testing datasets Set5\cite{Huang-CVPR-2015} and Set14\cite{Huang-CVPR-2015}. These two test sets has diverse set of images that truly signifies the performance of these models. As we can see from table III, our approach has achieved comparable performance to the others. Our proposed architecture achieves PSNR of 25.13 and SSIM of 0.794 on Set5\cite{Huang-CVPR-2015} and PSNR of 23.29 and SSIM of 0.701 on Set14\cite{Huang-CVPR-2015} which is very similar to other architectures while also maintaining real-time inference.

 Figure~\ref{fig:bird} depicts the comparison between  the high-resolution, low-resolution and our SwiftSRGAN's super-resolution output. The resultant super-resolution image is comparable to the high resolution image and reproduces the lighting and reflections entirely while also maintaining the color accuracy.

\section{Discussion}
    The proposed approach uses Depth-wise Separable Convolutions that has comparable performance with regular convolutions while also maintaining very low memory footprint and real-time inference. Depth-wise Separable Convolutions has very less number of trainable parameters compared to regular Convolutions. This reduces the size of the model and, number of computations during training and inference time. We used MobileNet instead of VGG in calculating loss which results in faster training time. While previously proposed approaches tend to focus more on the performance, we propose a real-time solution that can be run on low-end compute devices.

\section{Conclusion and Future Work}
    Image super-resolution is a challenging task and, many researchers have proposed architectures for this application which performs well requiring  very high computational capabilities and are not real-time. We propose a real-time efficient image super-resolution architecture that has comparable trade-off results with the previously proposed architectures and has the lowest inference time.
    
    Though this model achieves efficient real-time inference on low-end GPU, there is still scope for improving it's super-resolution performance. Our training data is relatively small when compared to other approaches that used 50,000 images from the ImageNet \cite{5206848} database for training. We are also planning to extend our work by improving our proposed network's performance.

\balance
\printbibliography

\end{document}